\documentclass{aa}  
\usepackage{graphicx}
\usepackage{natbib}
\bibpunct{(}{)}{;}{a}{}{,}
\usepackage{txfonts}
\newcounter{Rco}
\newcommand{\Ionst}[1]{\setcounter{Rco}{#1}\Roman{Rco}}
\newcommand{\Ion}[2]{\mbox{#1\,{\scriptsize\Ionst{#2}}}}

\newcommand{\loggw}[1]{\mbox{$\log g\hspace{-0.5mm} =\hspace{-0.5mm}  #1$}}
\newcommand{\ab}[1]{\mbox{Fig.\,\ref{#1}}}
\newcommand{\aba}[1]{\mbox{Figure\,\ref{#1}}}

\newcommand{\sA}[1]{\mbox{(Fig.\,\ref{#1})}}
\newcommand{\se}[1]{\mbox{Sect.\,\ref{#1}}}
\newcommand{\sK}[1]{\mbox{(Sect.\,\ref{#1})}}
\newcommand{\sla}{\raisebox{-0.10em}{$\stackrel{<}{{\mbox{\tiny $\sim$}}}$}}
\newcommand{\sle}{\mbox{\raisebox{0.20em}{{\tiny $\leq$}}}}
\newcommand{\sT}[1]{\mbox{(Table\,\ref{#1})}}
\newcommand{\ta}[1]{\mbox{Table\,\ref{#1}}}
\newcommand{\Teff}{\mbox{$T_\mathrm{eff}$}}
\newcommand{\TeffwkK}[1]{\mbox{$\Teff\hspace{-0.5mm} =\hspace{-0.5mm} #1\,\mathrm{kK}$}}
\newcommand{\Teffw}[1]{\mbox{$\Teff\hspace{-0.5mm} =\hspace{-0.5mm} #1\,\mathrm{MK}$}}
\newcommand{\exo}{EXO\,0748$-$676}

\newcommand{\kpd}{KPD\,0005+5106}
\newcommand{\lsv}{LS\,V\,$+46\degr 21$}

\newcommand{\vsgr}{V\,4743\,Sgr}
\begin{document}
\title{Absorption features in the spectra of X-ray bursting neutron stars}

\author{T\@. Rauch\inst{1}
   \and V\@. Suleimanov\inst{1, 2}
   \and K\@. Werner\inst{1}}

\institute{Institute for Astronomy and Astrophysics,
           Kepler Center for Astro and Particle Physics,
           Eberhard Karls University, 
           Sand 1,
           72076 T\"ubingen, 
           Germany,
           \email{rauch@astro.uni-tuebingen.de}
      \and
           Kazan State University, 
           Kremlevskaja Str\@., 18, 
           Kazan 420008,
           Russia
} 


\date{Received May 5, 2008; accepted August 22, 2008}

\authorrunning{T\@. Rauch et al.}

\abstract
{The discovery of photospheric absorption lines in
  \emph{XMM-Newton} spectra of
  the X-ray bursting neutron star in \exo\ by Cottam and collaborators allows us to constrain the
  neutron star mass-radius ratio from the measured gravitational
  redshift. A radius of $R=9-12$\,km for a plausible mass range
  of $M=1.4-1.8$\,M$_\odot$ was derived by these authors.}
{It has been claimed that the absorption features stem from gravitationally
redshifted ($z=0.35$) $n=2-3$ lines of H- and He-like iron. We investigate
this identification and search for alternatives.}
{We compute LTE and non-LTE neutron-star model atmospheres and detailed synthetic
spectra for a wide range of effective temperatures (\Teffw{1-20})
and different chemical compositions.}
{We are unable to confirm the identification of the absorption features in the
X-ray spectrum of \exo\ as $n=2-3$ lines of H- and He-like iron
(\ion{Fe}{xxvi} and \ion{Fe}{xxv}). These are subordinate
lines that are predicted by our models to be too weak at any \Teff. It is
more likely that the strongest feature is from the $n=2-3$ resonance transition in
\ion{Fe}{xxiv} with a redshift of $z=0.24$. Adopting this value yields a
larger neutron star radius, namely $R=12-15$\,km for the mass range $M=1.4-1.8$\,M$_\odot$, favoring a stiff
equation-of-state and excluding mass-radius relations based on exotic
matter. Combined with an estimate of the stellar radius $R > 12.5$\,km from the work of
\"Ozel and collaborators, the $z=0.24$ value provides a minimum neutron-star mass
of $M > 1.48$\,M$_\odot$, 
instead of $M > 1.9$\,M$_\odot$, when assuming $z=0.35$.}
{The current state of line identifications in the neutron star of \exo\
must be regarded as highly uncertain. Our model atmospheres show that
lines other than those previously thought must be associated with the observed
absorption features.}


\maketitle
%

\section{Introduction}
\label{sect:intro}

\citet[][hereafter CPM02]{cpm2002} identified discrete absorption
features corresponding to electronic transitions in highly ionized iron
in the burst spectra of the neutron star in \exo\ observed with
\emph{XMM-Newton}. They identified $n=2-3$ absorption features of H-like
(\ion{Fe}{xxvi}) and He-like iron (\ion{Fe}{xxv}) in early and late
phases of the bursts, respectively. These features provided a redshift measurement of
$z=0.35$, corresponding to a mass-radius ratio of
$M/R=0.152$\,M$_\odot$\,km$^{-1}$. Using this redshift and an estimate
for the stellar radius of $R>13.8$\,km, \citet{ozel06} inferred a
neutron star mass of $M>2.10$\,M$_\odot$.
However, using \"Ozel's numbers and formulae, we obtain slightly
different values ($R>12.5$\,km, $M>1.9$\,M$_\odot$).

The identification of only a few observed lines in burst spectra of NS
with unknown redshift is potentially ambiguous. It is the aim of our
paper to confirm the proposed line identifications
in \exo. To this end, we performed LTE and non-LTE model-atmosphere
calculations in a wide parameter range using two independently developed stellar atmosphere
modeling codes. In this systematic study, we elaborate on our
earlier suspicion that an alternative line identification is more likely
\citep{wea2007}.

In the past two decades, spectral analysis of hot, compact stars by
means of fully line-blanketed NLTE model atmospheres
\citep[e.g.][]{r2003} has achieved a high level of
sophistication. For our analyses, the T\"ubingen NLTE Model Atmosphere
Package
\citep[\emph{TMAP}\footnote{http://astro.uni-tuebingen.de/$^\sim$rauch/TMAP/TMAP.html},][]{wea2003,
rd2003} was used to calculate plane-parallel NLTE model atmospheres
that are in radiative and hydrostatic equilibrium.  Such model
atmospheres were used successfully in the analysis of hot white
dwarfs, e.g\@.  \lsv\ \citep[\TeffwkK{95},][]{rea2007} and  \kpd\
\citep[\TeffwkK{200},][]{wrk2008}.  \emph{TMAP} models were also
calculated for the extremely hot super-soft X-ray source \vsgr\
\citep[\TeffwkK{610},][]{rea2005}.

The \emph{TMAP} NLTE model atmospheres can also be employed in the
analysis of neutron stars with low magnetic fields,  i.e\@. in the range
where the magnetic field strength has no significant impact on atomic
data  ($B \sla 10^{12}\,\mathrm{G}$). Since magnetic fields in low-mass
X-ray binaries (LMXBs) are believed to be small, X-ray spectra of the
neutron star in \exo\ can be compared with our synthetic spectra.

We calculated \emph{TMAP} models for the relevant \Teff\ range and
investigated their \Teff-dependence \sK{sect:models}. We describe
results of a comparison of LTE and NLTE  model-atmosphere fluxes in
\se{sect:LTEvsNLTE}. A comparison of \exo\ X-ray observations with our
models follows in \se{sect:exo0748-676} and we conclude in
\se{sect:conclusions}.

\section{Model atmospheres and atomic data}
\label{sect:models}

The \emph{TMAP} code was not especially designed specifically for calculating the
burst spectra of neutron stars in LMXBs. \emph{TMAP} does not consider
general relativistic effects on the radiation transfer within the
atmosphere, or velocity fields. Nevertheless, we believe that \emph{TMAP}
models are well suited for our purpose, which is line
identification in observed spectra.  Comptonization effects are also
neglected because \citet{sw2007} demonstrated that at \Teffw{3} and
\loggw{14.3} its impact on the emergent spectrum is detectable only at
energies higher that about 10\,keV ($\la 1.24$\,\AA).  It is
clear that Comptonization is more important for models with higher
\Teff\ and that it determines the temperature structures and emergent spectra of
the hottest models  \citep[$T_{\rm eff} \ga 20\,\mathrm{MK}$,
see][]{Lapidusetal:86,Londonetal:86,Madej:91,Madej.etal:04}.  The
influence of Comptonization is discussed in detail in the next
section.

Throughout the paper, we fix the surface gravity in all models to be \loggw{14.39}
(cm/sec$^2$) representing a neutron star with $M =
1.4\,\mathrm{M_\odot}$ and  $R = 10\,\mathrm{km}$.

\begin{figure}[ht!]
  \resizebox{0.98\hsize}{!}{\includegraphics{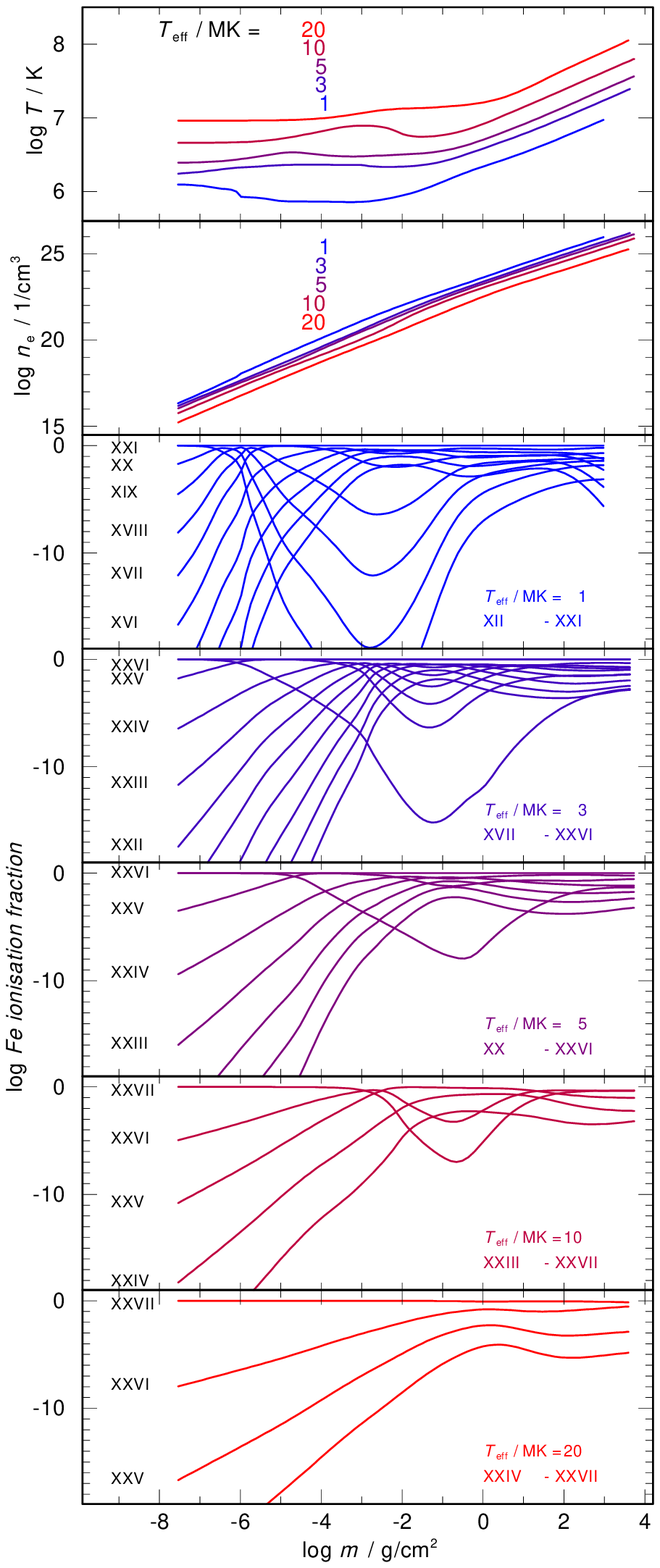}}
  \caption[]{Atmosphere structure 
            (temperature: 1$^\mathrm{st}$ panel from top, 
            electron density: 2$^\mathrm{nd}$)
            of our NLTE models which consider opacities of H, He, C, N,O, and Fe with
            solar abundances, \Teffw{1, 3, 5, 10, 20}. 
            In panels 3 to 7, the ionization fractions of the iron ions which are used 
            in the calculations are shown. 
            Depending on \Teff, the considered ions are selected to well represent the
            dominant ionization stages in the line-forming region.
            }
  \label{fig:structure}
\end{figure}

The model atoms used in our NLTE calculations are summarized in
\ta{tab:statistics}. From test calculations, we determined the
dominant ionization stages and adjusted the model atoms \sT{tab:femodel}
to avoid numerical instabilities due to extremely depopulated
ionisation stages. \aba{fig:structure} displays the atmospheric
structures for models with different \Teff.  For all elements, we
consider level dissolution (pressure ionization) by following
\citet{hm1988} and \citet{Lanz.Hub:94}. 
In the latter paper, the \citet{hm1988} method was also considered for the hydrogen
atom, but it can be used for other ions as well. The extended 
opacity tables were indeed calculated in the framework of the Opacity Project 
\citep{OP} up to temperatures of $\sim 10^7\,\mathrm{K}$ and densities of $\sim 10^2\,\mathrm{g\,cm}^{-3}$. 
These opacities agree well with the opacities 
calculated by the OPAL project \citep{OPAL}. In the OPAL project, the other 
``physical method'' for the occupation densities calculation is used.
The emergent spectra in our models are formed at densities of $\le$ 10
g cm$^{-3}$, therefore our calculations of the occupation densities are correct.
Because of the high particle
density in the neutron star atmospheres, this is of course an important
point for the correct computation of the atmospheric structure in deeper
layers.  As an example, we display the occupation probabilities of
the energy levels of \ion{Fe}{xxiv} in a particular model
\sA{fig:FE24_HM}. Level dissolution (i.e\@. low occupation probability) is
significant for all atomic levels.

\begin{table}[ht!]
\caption{Statistics of model atoms used in the calculation
         of our \emph{TMAP} models. (N)L is the number of levels
         treated in (N)LTE, RBB (radiative bound-bound)
         is the number of line transitions.
         H, He, C, N, and O are the same for all \Teff,
         while the Fe ionization stages, which are considered,
         are adjusted, respectively \sT{tab:femodel}.
         } 
\label{tab:statistics}
\small{
\begin{tabular}{p{8mm}cccp{11mm}ccc}
\hline
\hline
\noalign{\smallskip}
ion & NL & L & RBB & Fe ion & NL & L & RBB \\
\hline
\noalign{\smallskip}
\Ion{H }{1} &  10 &  6 & 45 & \Ion{Fe}{12} &   5 &  1 &  3 \\
\Ion{H }{2} &   1 &  - &  - & \Ion{Fe}{13} &   7 &  4 &  3 \\
\Ion{He}{2} &  16 & 18 & 95 & \Ion{Fe}{14} &   8 &  4 &  8 \\
\Ion{He}{3} &   1 &  - &  - & \Ion{Fe}{15} &   7 &  6 &  5 \\
\Ion{C }{5} &  29 & 21 & 60 & \Ion{Fe}{16} &   7 &  5 & 10 \\
\Ion{C }{6} &  21 & 15 & 49 & \Ion{Fe}{17} &   6 &  8 &  4 \\
\Ion{C }{7} &   1 &  - &  - & \Ion{Fe}{18} &   4 &  4 &  2 \\
\Ion{N }{6} &  17 & 36 & 33 & \Ion{Fe}{19} &   5 &  1 &  3 \\
\Ion{N }{7} &  21 & 34 & 55 & \Ion{Fe}{20} &   6 &  2 &  4 \\
\Ion{N }{8} &   1 &  - &  - & \Ion{Fe}{21} &   9 &  4 &  6 \\
\Ion{O }{7} &  19 & 16 & 33 & \Ion{Fe}{22} &   6 &  3 &  4 \\
\Ion{O }{8} &  15 & 30 & 30 & \Ion{Fe}{23} &   7 &  8 &  2 \\
\Ion{O }{9} &   1 &  - &  - & \Ion{Fe}{24} &  12 & 15 & 28 \\
            &     &    &    & \Ion{Fe}{25} &  23 & 30 & 59 \\
            &     &    &    & \Ion{Fe}{26} &  15 & 40 & 30 \\
            &     &    &    & \Ion{Fe}{27} &   1 &  - &  - \\
\hline 
\end{tabular} 
}
\end{table}

\begin{table}[ht!]
\caption{Fe ionization stages that are considered at individual \Teff\ (cf\@. \ta{tab:statistics} and \ab{fig:structure}).
         } 
\label{tab:femodel}
\begin{tabular}{r@{\,-\,}lr@{\,\,\,-}rp{2mm}r@{\,-\,}lr@{\,\,\,-}r}
\hline
\hline
\noalign{\smallskip}
\multicolumn{2}{c}{\Teff\ / MK} & \multicolumn{2}{c}{ionization stages} &&
\multicolumn{2}{c}{\Teff\ / MK} & \multicolumn{2}{c}{ionization stages} \\
\hline
\noalign{\smallskip}
\multicolumn{2}{c}{1} & {\sc xii}   & {\sc xxi}  && \multicolumn{2}{c}{6} & {\sc xxi}   & {\sc xxvii} \\
\multicolumn{2}{c}{2} & {\sc xvi}   & {\sc xxv}  && \multicolumn{2}{c}{7} & {\sc xxii}  & {\sc xxvii} \\
\multicolumn{2}{c}{3} & {\sc xvii}  & {\sc xxvi} &&                8 & 13 & {\sc xxiii} & {\sc xxvii} \\
\multicolumn{2}{c}{4} & {\sc xix}   & {\sc xxvi} &&               14 & 20 & {\sc xxiv}  & {\sc xxvii} \\
\multicolumn{2}{c}{5} & {\sc xx}    & {\sc xxvi} && \multicolumn{4}{c}{}                              \\
\hline 
\end{tabular} 
\end{table}

\begin{figure}[ht!]
  \resizebox{\hsize}{!}{\includegraphics{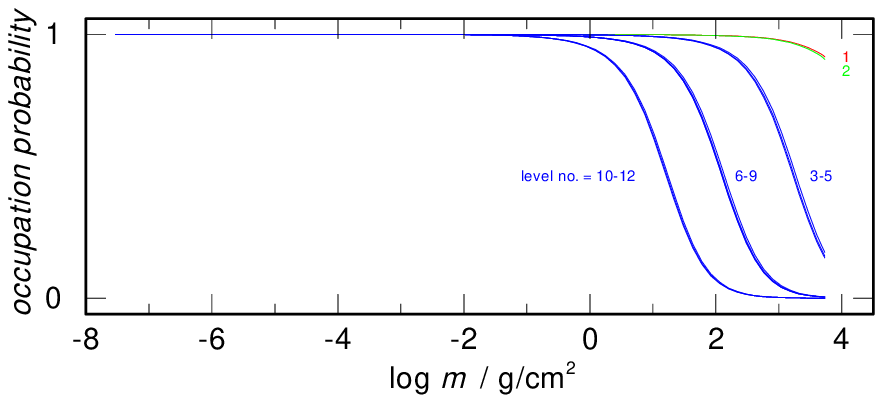}}
  \caption[]{Occupation probabilities of the \ion{Fe}{xxiv} levels
    (labeled consecutively with increasing excitation energy) in
    our NLTE model atmosphere with \Teffw{10} and solar abundances.
            }
  \label{fig:FE24_HM}
\end{figure}

The model atoms are constructed using energy levels from 
NIST\footnote{http://physics.nist.gov/PhysRefData/ASD/index.html} (National Institute of Standards and Technology)
and oscillator strengths and photoionization cross-sections calculated by the Opacity Project 
(TIPTOPbase\footnote{http://vizier.u-strasbg.fr/topbase/}).
The complete set of model ions used is available from
\emph{TMAD}, the T\"ubingen Model-Atom Database\footnote{http://astro.uni-tuebingen.de/$^\sim$rauch/TMAD/TMAD.html}.

We compute plane-parallel non-LTE model atmospheres in hydrostatic and
radiative equilibrium.
Our calculations start from grey model atmospheres that are calculated by \emph{TMAP} for the range of parameters
$-8.0 \leq \log \tau_{\rm Ross} \leq +2.6$. The atmospheres are represented by 90 depth points, set up 
equidistantly in $\log \tau_{\rm Ross}$ between points 1 and 85 (outside to inside), decrease logarithmically
by a factor of two from point to point towards the inner
boundary. The subsequent non-LTE modeling is performed after
transforming from the $\tau_{\rm Ross}$ on a column-mass
scale $m$.

Line broadening due to the Stark effect is considered.
According to \citet{c71}, the broadening due to the quadratic Stark effect
is approximately given by:

\begin{equation}
\gamma_{Stark} = 5.5\times10^{-5}\;\frac{n_e}{\sqrt{T}}\;\left
    [\,\frac{(n_{\rm eff}^{\rm up})^2}{z+1}\,\right]^2
\end{equation}

where $n_{\rm eff}^{\rm up}$ is the effective principal quantum number of the
upper level, and $z$ is the effective charge seen by the active electron.

The linear Stark effect (e.g\@. in the case of \Ion{Fe}{26}) 
is considered using an approximate formula 
\citep{u68,whh91}

\begin{equation}
\kappa(\Delta\lambda) = \frac{\pi e^2}{m c^2}\lambda^2 f \frac{1}{s^*_n F_0}
U\left(\frac{\Delta\lambda}{s^*_n F_0}\right)   
\end{equation}

with the electric microfield 

\begin{equation}
F_0 = 2.61 e\,\left[\,\sum_{\rm ions }^{~}z_i^{3/2}n_i\,\right]^{2/3}
\end{equation}

where $U(\beta)$ is given by \citet{vd49}.
A measure of the width of the Stark pattern is given by

\begin{equation}
s_n=0.0192\lambda^2\left[\,n_{\rm up}(n_{\rm up}-1)+n_{\rm low}(n_{\rm low}-1)\,\right]/Z .
\end{equation}

The dependence of the emergent flux on \Teff\ is illustrated in \ab{fig:fluxes}. 
We note that lines of \ion{C}{vi}, \ion{N}{vii}, and \ion{O}{viii} are detectable in the synthetic spectra
up to \Teff$\approx 12\,\mathrm{MK}$. 
Lines of \ion{Fe}{xxiii} -- \ion{Fe}{xxvi}
visible in the wavelength range from 7 to 12\,\AA\ (no gravitational redshift considered) 
show a strong dependence on \Teff\ and are therefore very sensitive temperature indicators.

\begin{figure}[ht!]
  \resizebox{0.96\hsize}{!}{\includegraphics{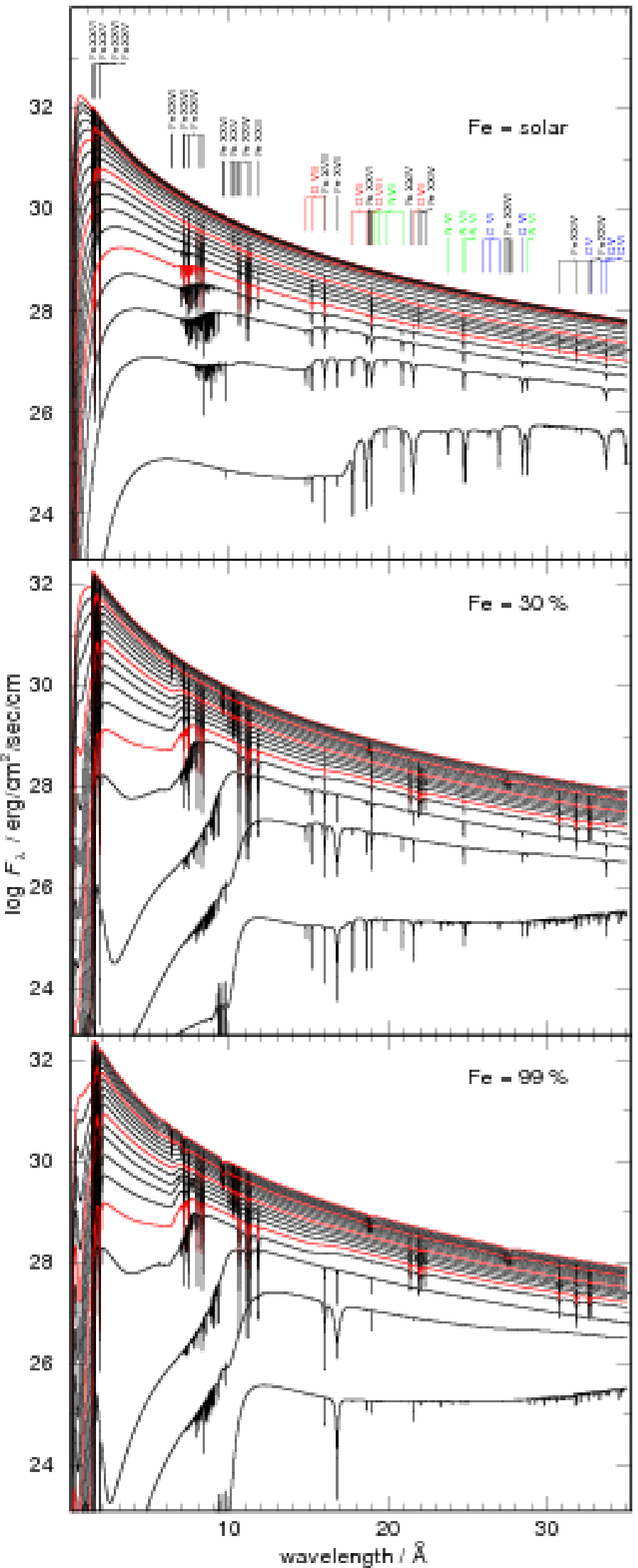}}\vspace{-1mm}
  \caption[]{Astrophysical fluxes of our NLTE models. The three panels
    display models with different chemical composition (solar,
             33\,\% (by mass),  and 99\,\% Fe content). In each panel
             models with different effective temperatures are shown
             (\Teffw{1 - 20}, from bottom to top). Line identifications
             are given in the top panel. 
            }
  \label{fig:fluxes}
\end{figure}

The iron abundance in the photosphere has a significant influence on the
emergent spectrum.  In \ab{fig:fluxes}, we show fluxes calculated from
model atmospheres with solar, 30\%, and 99\% iron abundance (by mass; H,
He, C, N, and O are considered with solar abundance ratios relative to
each other).

In the framework of the Virtual
Observatory\footnote{http://www.ivoa.net} (\emph{VO}), all spectral
energy distributions (SEDs, $\lambda - F_\lambda$)  from the \emph{TMAP}
model grids described here are available in \emph{VO} compliant form
from the \emph{VO} service  \emph{TheoSSA}
\footnote{http://vo.ari.uni-heidelberg.de/ssatr-0.01/TrSpectra.jsp?}
provided by the \emph{German Astrophysical Virtual Observatory}
(\emph{GAVO}\footnote{http://www.g-vo.org}) as well as
atables\footnote{http://astro.uni-tuebingen.de/$^\sim$rauch/TMAF/TMAF.html}
for the use with
\emph{XSPEC}\footnote{http://heasarc.gsfc.nasa.gov/docs/xanadu/xspec}.

\section{LTE versus NLTE modeling}
\label{sect:LTEvsNLTE}

For the ground states of iron ions (\ab{fig:departure}, upper panel) in a particular 
model atmosphere  (\Teffw{10}), the necessity of considering NLTE effects 
appears obvious from the LTE departure coefficients, which show the ratio 
of LTE and NLTE occupation numbers, where the LTE population number is 
defined relative to the ground state of the subsequent (next highest) ionization stage.
However, a closer look into the model
structure shows that all atomic levels connected by the line features discussed
in the context of \exo\ have departure coefficients close to
unity. The relative deviation from the LTE population density is at
most 10\% in the line-forming region. For the $n=2-3$ lines of
\ion{Fe}{xxiv-xxvi}, this region is confined to the parameter range of $-1.2 \le \log m \le +0.2$. 
As a consequence, NLTE effects on the line profiles are
expected to be small. 
A direct comparison of the relative densities of the iron ions in LTE and 
NLTE models confirms this conclusion (\ab{fig:departure}, bottom panel; see also below).

\begin{figure}[ht!]
  \resizebox{\hsize}{!}{\includegraphics{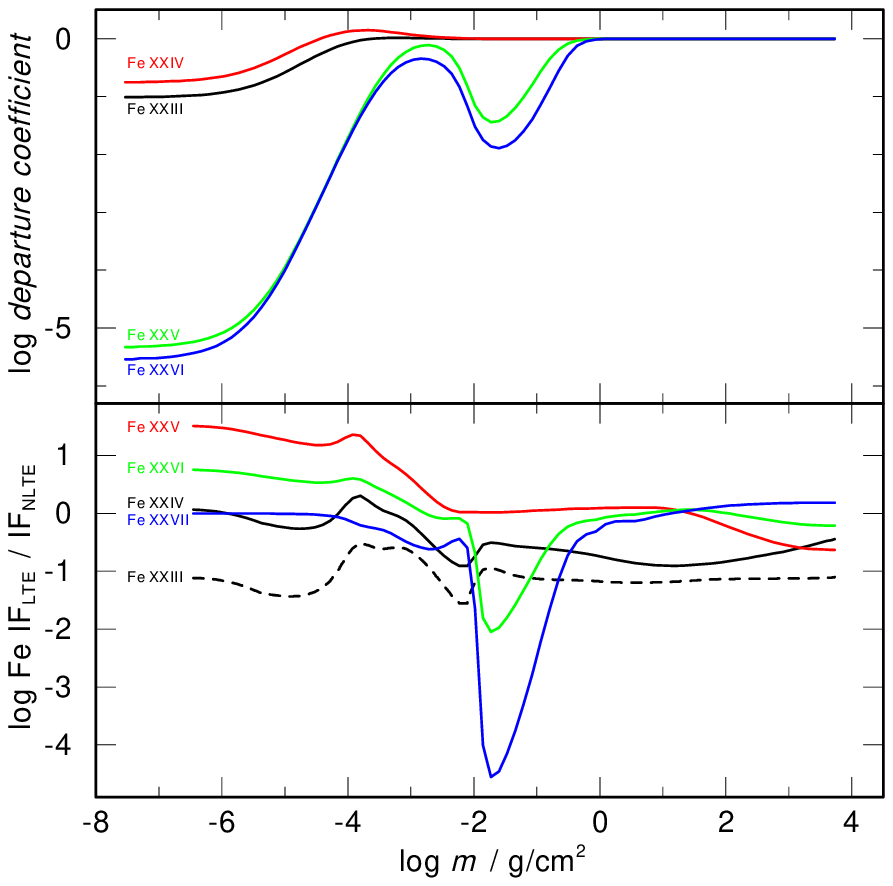}}
  \caption[]{{\it Upper panel:} Departure coefficients of the Fe\,{\sc xxiii} -- {\sc xxvi} ground states in our 
\Teffw{10} model atmosphere (solar abundances). {\it Bottom panel:} Ratios of ionisation fractions (IF)
of the iron ions for the LTE and NLTE model atmospheres with identical parameters. 
            }
  \label{fig:departure}
\end{figure}

For a more detailed comparison, we calculated LTE models
(cf\@. \citet{sw2007} and  \citet{Ibragimov.etal:03}). In the LTE model-atmosphere
code, we are able to model the opacities due to the photoionization of all ions 
of the 15 most abundant chemical elements, and about 25\,000 spectral lines 
of all ions of 26 elements (all elements with
atomic number $A \leq 30$, apart from F, Sc, V, and Cu).
Atomic line data were taken  from the CHIANTI database
\citep{dere97}. As for the NLTE atmospheres, we used the
occupation-probability formalism  for ion population calculations for all
considered elements. Compton scattering was also taken into account with
this LTE code \citep{Sul.Pout:06, sw2007}.

\begin{figure}[ht!]
  \resizebox{\hsize}{!}{\includegraphics{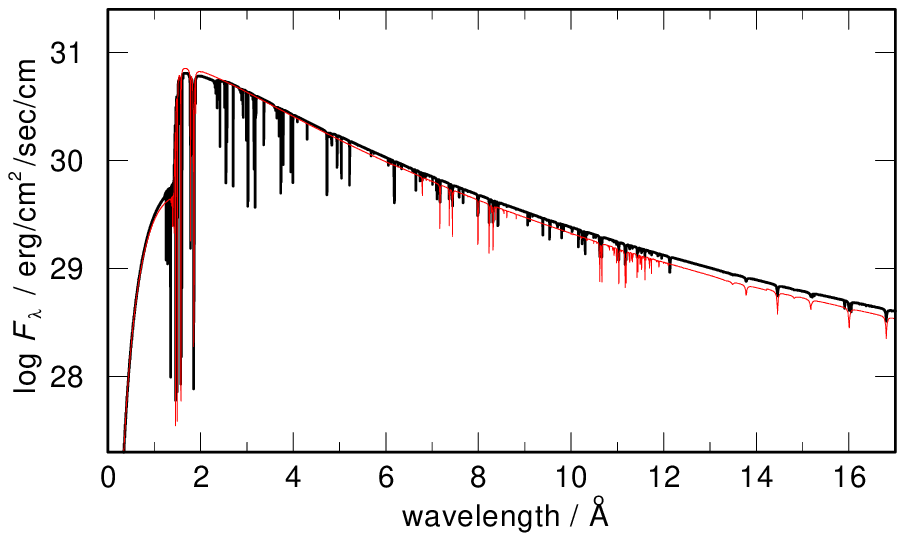}}
  \resizebox{\hsize}{!}{\includegraphics{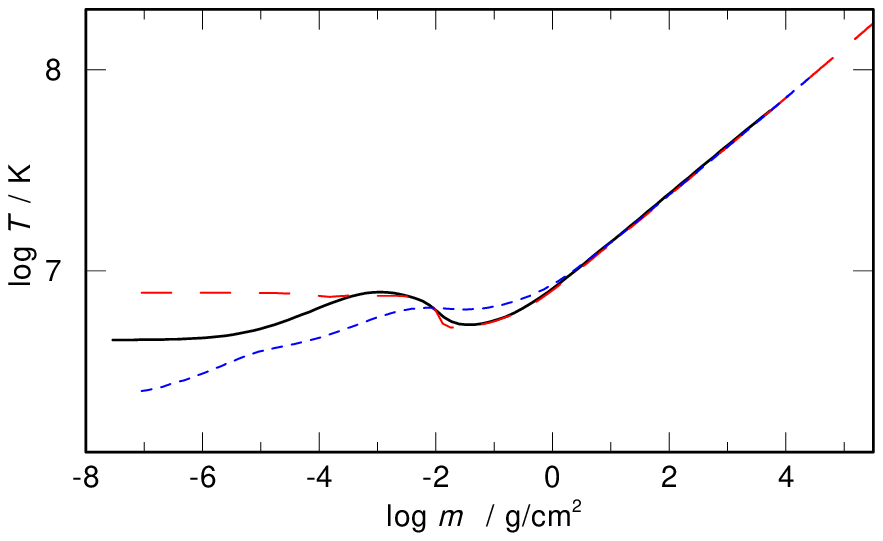}}
  \caption[]{{\it Top panel:} emergent (unredshifted) spectra of LTE models with 
              \Teffw{10} and different chemical compositions.
              Thick line:  model of solar composition for all chemical
              elements up to $Z=30$, 
               thin line: model with H, He, C, N, O, Fe only (solar abundances). 
             {\it Bottom panel:} temperature structures of model atmospheres with \Teffw{10}. 
               Solid lines: NLTE model with H, He, C, N, O, Fe only (solar abundances), 
               long dashes:  LTE model with the same chemical
              composition (note that the high-temperature plateau in
              this model is caused by Compton scattering), 
              short dashes:  LTE model with solar composition for all chemical elements.
            }
  \label{fig:sul_f1}
\end{figure}

\begin{figure}[ht!]
  \resizebox{\hsize}{!}{\includegraphics{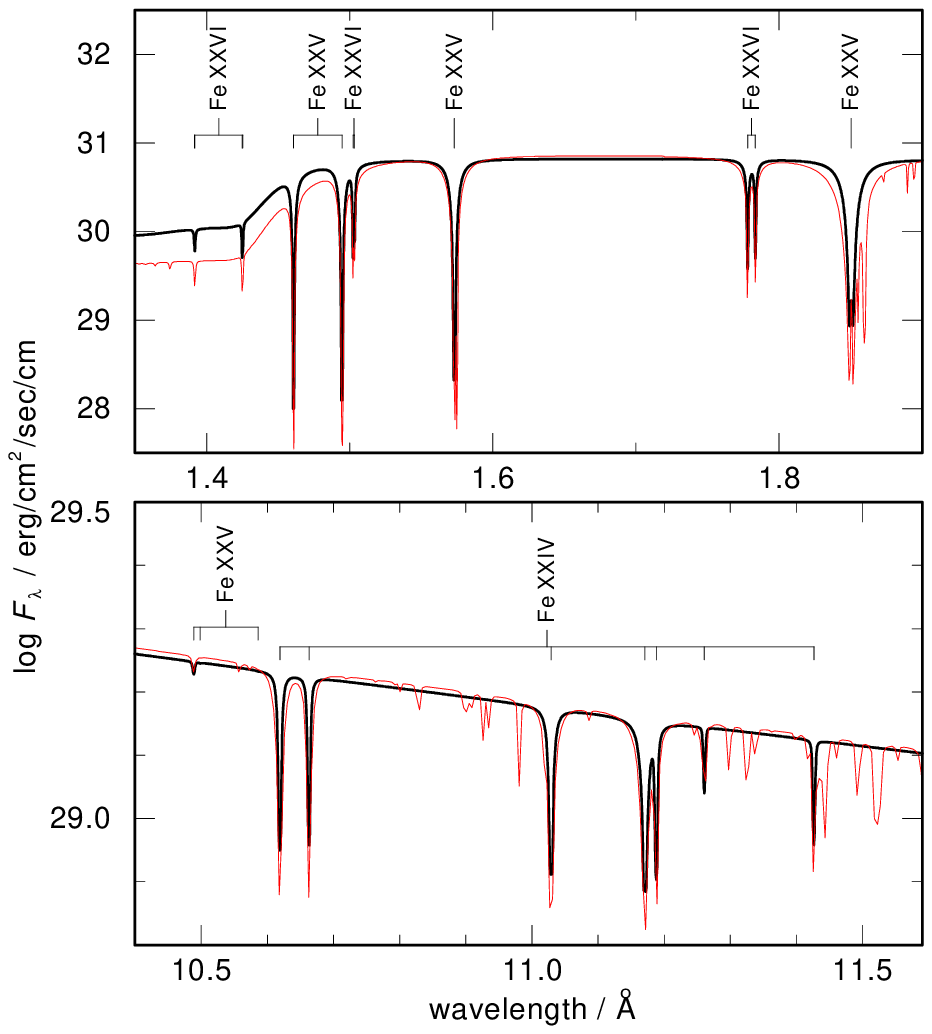}}
  \caption[]{Comparison of the emergent (unredshifted) spectra of NLTE (thick lines) and LTE models (thin)
             with \Teffw{10} and identical (solar) chemical composition in two different spectral bands.
             Lines considered in both models are marked.
            }
  \label{fig:sul_f2}
\end{figure}

A comparison of LTE models with one of the NLTE models is
shown in Figs.~\ref{fig:sul_f1} and \ref{fig:sul_f2}. We computed three
LTE models of the same effective temperature and surface gravity
(\Teffw{10}, \loggw{14.39}) but with different chemical compositions. The
first model had the same chemical composition as the NLTE model (H,
He, C, N, O, and Fe all of solar abundances). The second model had solar
chemical composition for all considered elements ($A \leq 30$), and the
third model was calculated without hydrogen. It is well known
\citep{lpt93} that on a neutron-star surface, the accreted hydrogen can
transform into helium by means of pycnonuclear reactions (depending on
the accretion rate), and part of the X-ray bursting neutron stars have
helium  envelopes without hydrogen. Therefore, we considered a helium
model with solar composition of the other chemical elements.

The temperature structure of the first model was similar to the
temperature structure of the NLTE model at $\log m > -3$, but the
temperature structure of the second model differed in the
layers with $\log m < 0$, where the emergent flux was formed
(\ab{fig:sul_f1}, bottom panel). Therefore, the chemical composition of
the heavy elements was more important than the NLTE effects. This was
also true for spectral energy distribution (\ab{fig:sul_f1}, top panel;
\ab{fig:sul_f2}). This was due to the influence of the large number of
other heavy element spectral lines in the 2 -- 8\,\AA\ band (compare
spectra in Figs\@. \ref{fig:fluxes} and \ref{fig:sul_f1}). The spectrum of the
first model was similar to that of the NLTE model than the spectrum of 
the second model (Figs\@. \ref{fig:sul_f1}
and \ref{fig:sul_f2}).  The difference between the emergent spectra of
NLTE  and LTE model atmospheres with identical chemical composition is
significant only at $\lambda < 2$\,\AA.  The spectral line
strengths are weaker in the second model because the temperature is higher
in the line-formation layers ($-2.6 < \log m < +0.4$)
(\ab{fig:sul_f1}, bottom panel; \ab{fig:sul_f5}).

\begin{figure}[ht!]
  \resizebox{\hsize}{!}{\includegraphics{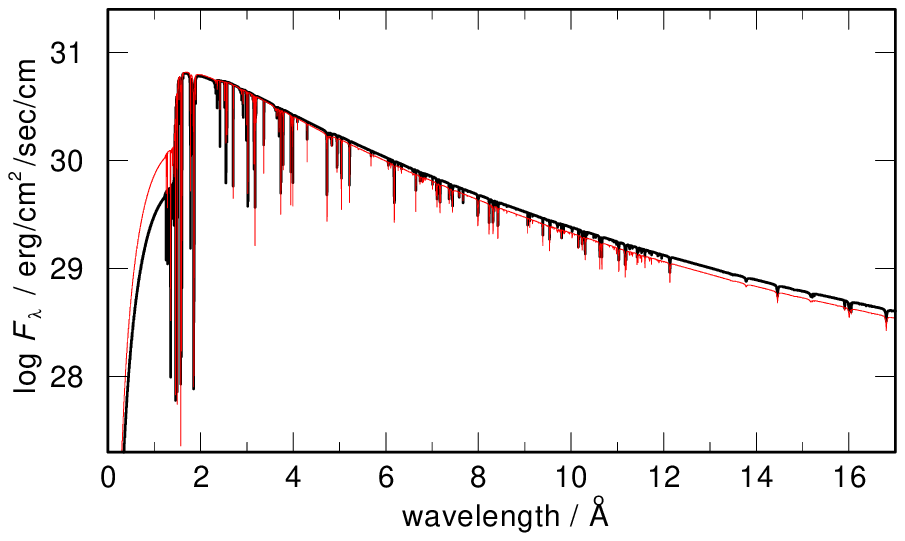}}
  \resizebox{\hsize}{!}{\includegraphics{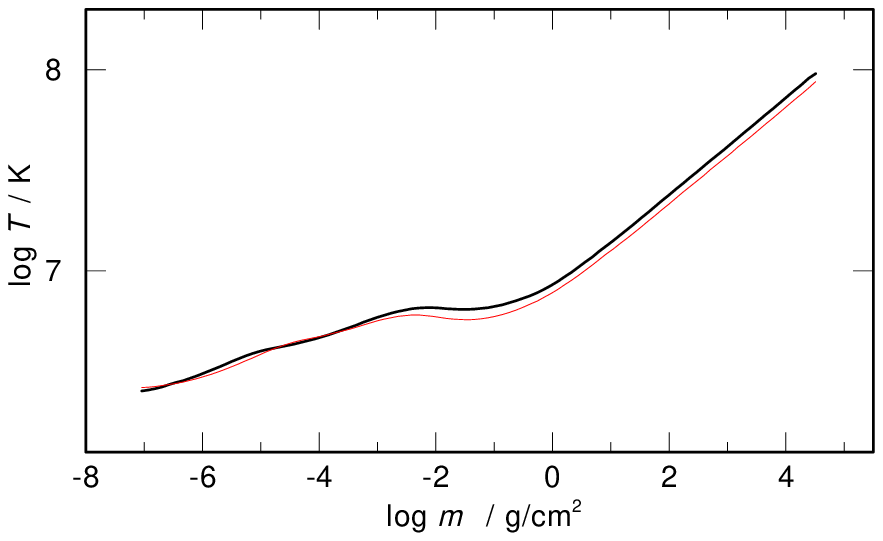}}
  \caption[]{Emergent (unredshifted) spectra (top panel) and temperature structures (bottom panel) of 
             LTE model atmospheres with \Teffw{10} and different chemical compositions. 
              Thick line: model with solar composition for all chemical elements, 
                    thin: helium model (without hydrogen) with solar abundance of heavy elements.}
  \label{fig:sul_f3}
\end{figure}

The helium model atmosphere had lower temperatures at deep layers ($\log
m > -2$) than the second model, and it showed slightly
stronger depths of the spectral lines (Figs\@. \ref{fig:sul_f3} and
\ref{fig:sul_f5}).  However, the difference between helium and hydrogen
models was not significant, and the qualitative conclusions  for the
hydrogen models were also applicable to the helium models.

\begin{figure}[ht!]
  \resizebox{\hsize}{!}{\includegraphics{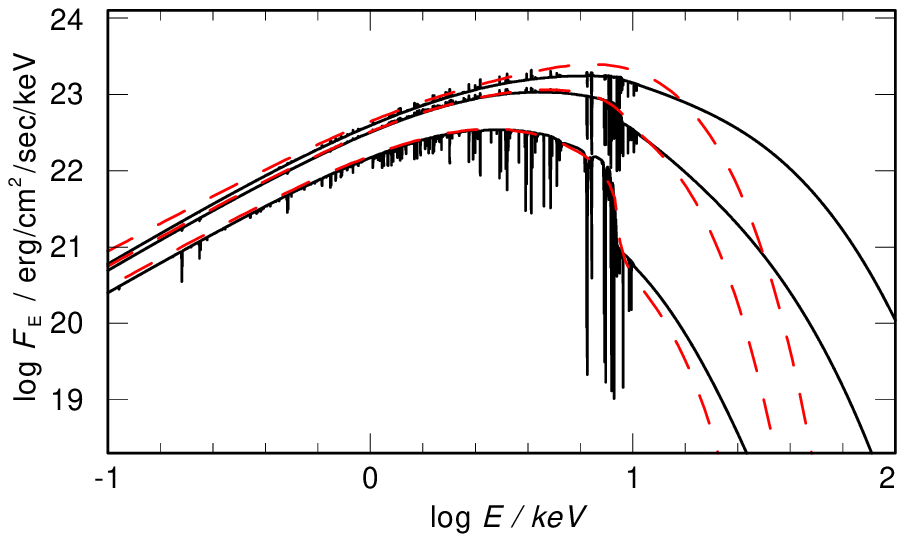}}
  \resizebox{\hsize}{!}{\includegraphics{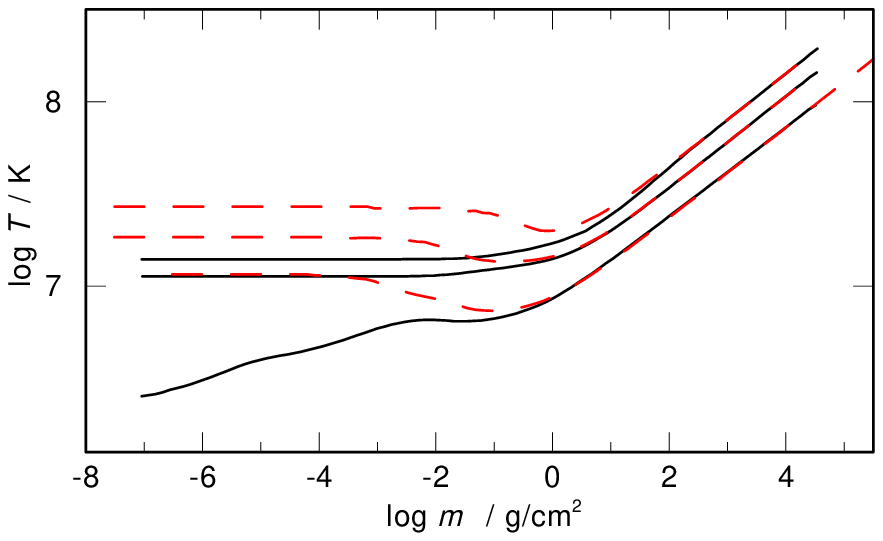}}
  \caption[]{Emergent (unredshifted) spectra (top panel) and temperature structures (bottom panel) of 
             LTE models with \Teffw{10, 15, 20} for Thomson (solid curves) and 
             Compton scattering (dashed curves).
            }
  \label{fig:sul_f4}
\end{figure}

LTE model atmospheres of solar composition were also calculated by taking
Compton scattering into account.  In \ab{fig:sul_f4}, the temperature
structures and emergent spectra of these models compared with
Thomson-scattering models are shown. It is possible to make the
following conclusion from this figure.  Comptonization is
significant for models with $\Teff \ge  15\,\mathrm{MK}$, but the models
with lower  temperatures and solar chemical composition can be
considered without Compton scattering.

\section{\exo}
\label{sect:exo0748-676}

\begin{figure*}[ht!]
  \resizebox{\hsize}{!}{\includegraphics{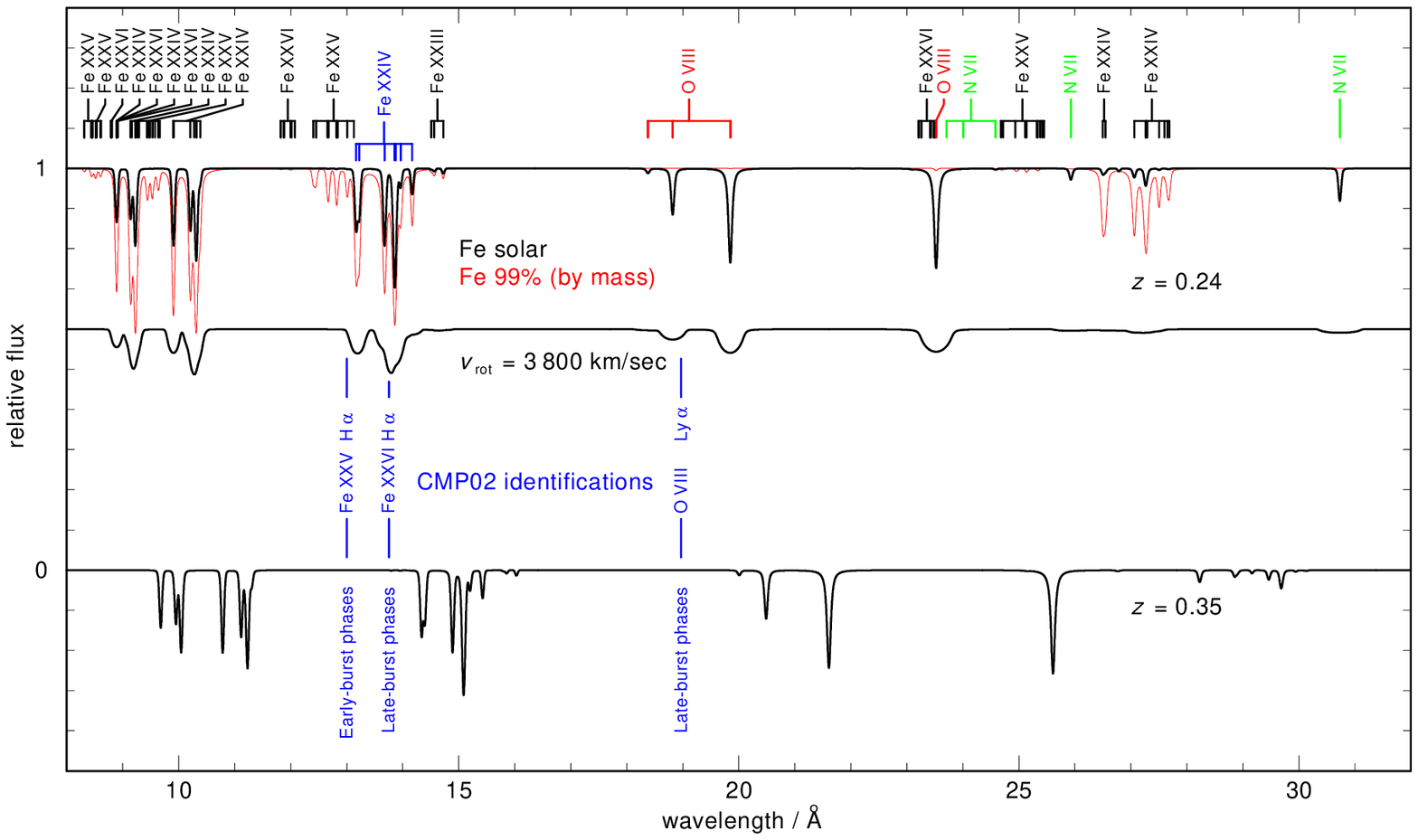}}
  \caption[]{Comparison of NLTE model-atmosphere fluxes calculated from
             \Teffw{8} models with different iron content. This figure can be
             directly compared with the observed spectra of \exo\
             presented in Fig.\,1 of CPM02 (therefore, all spectra are convolved with a Gaussian of FWHM = 0.124\,\AA, 
             which is about 2.5 times lower in quality than \emph{XMM-Newton's} RGS spectral resolution).
             At $z=0.35$ (bottom), there is no fit to the observation. 
             The identification of \ion{Fe}{xxv} by CMP02 is unlikely. 
             At a significantly lower redshift of $z=0.24$ (top), 
             the observed line features would originate instead from \ion{Fe}{xxiv}. 
              
            }
  \label{fig:nature}
\end{figure*}

\begin{figure}[ht!]
  \resizebox{\hsize}{!}{\includegraphics{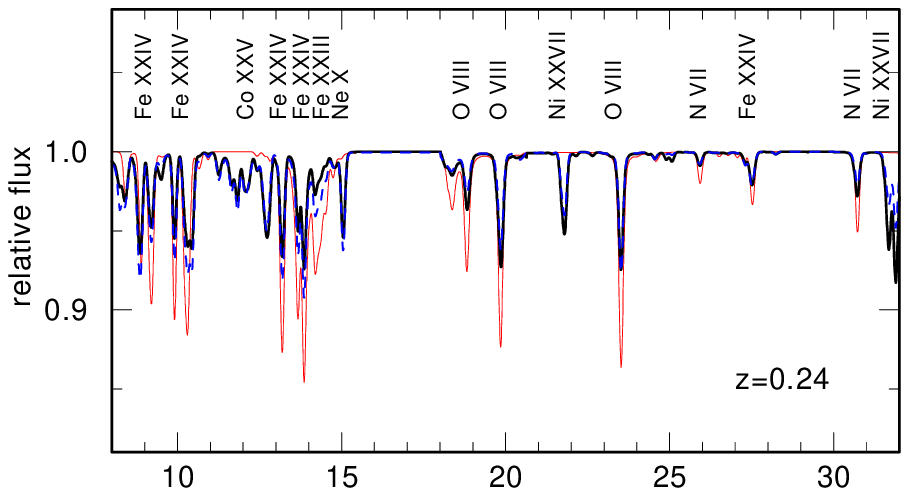}}
  \resizebox{\hsize}{!}{\includegraphics{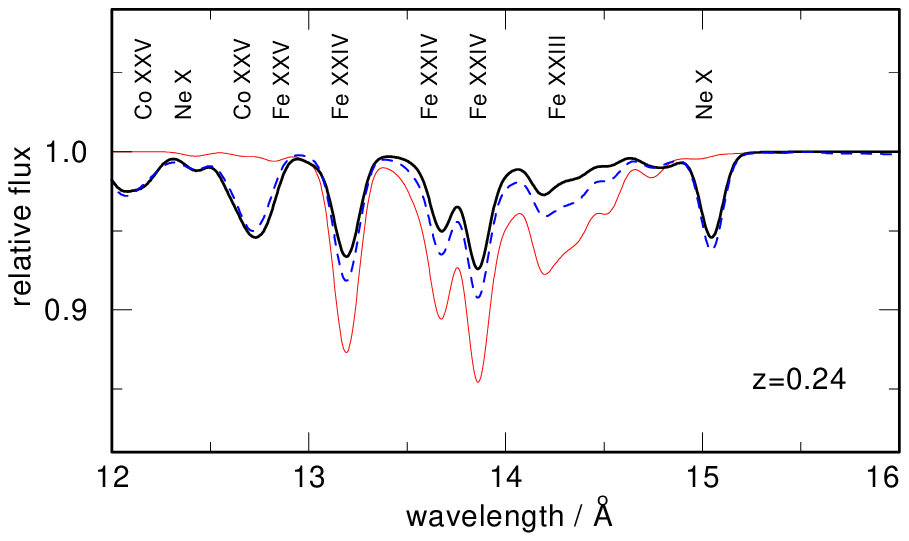}}
  \caption[]{Emergent redshifted ($z$ = 0.24), normalized fluxes of our LTE models with 
             \Teffw{10} and different chemical compositions. 
             Thick line: model with solar composition for all chemical elements, 
             dashed: model with H, He, C, N, O, Fe only (solar abundances), 
             thin: helium model. 
             All fluxes have been convolved with a Gaussian (FWHM = 0.124\,\AA, see \ab{fig:nature}).
            
            }
  \label{fig:sul_f5}
\end{figure}

CPM02 used blackbody flux distributions to determine the \Teff\ of the
neutron star in \exo\ in the early- and late-burst phase and derived
$k_\mathrm{B}\Teff \approx 1.8\,\mathrm{keV}$ ($\Teff \approx
20.9\,\mathrm{MK}$) and  $k_\mathrm{B}\Teff \sle 1.5\,\mathrm{keV}$
($\Teff \sle 17.4\,\mathrm{MK}$), respectively.  They identified
photospheric absorption lines of \ion{Fe}{xxv} at a redshift of
$z=0.35$.  In Figs\@. \ref{fig:nature} and \ref{fig:sul_f5}, we compare
theoretical spectra for the \emph{XMM-Newton} wavelength range.  It is
well known \citep{Lapidusetal:86,Londonetal:86, Madej:91, Madej.etal:04}
that the color temperature of an X-ray bursting neutron stars is higher
than the effective temperature by a factor of $1.5-1.7$; for comparison with
late-burst phase observations, we therefore
chose models with $T_{\rm eff} \approx$ 10\,MK.

\aba{fig:nature} (top) displays spectra from our \Teffw{8}
models in the wavelength range $8-32$\,\AA. It can be compared directly
with the observed spectra of \exo\ shown in Fig.\,1 of CPM02.  
To consider rotational broadening, we assume that $R=13.5\,\mathrm{km}$ (cf\@.
\ab{fig:eos}). At $\nu_\mathrm{rot} = 45\,\mathrm{Hz}$
\citep{vs2004}, this results in an equatorial velocity
$v_\mathrm{rot}=3\,800\,\mathrm{km/sec}$. The
synthetic spectrum in the middle panel of
\ab{fig:nature} is convolved 
with the respective rotational profile (see \citet{cea2006}
for a more detailed treatment of rotational broadening).

The \ion{Fe}{xxv} $n=2-3$ absorption features (at rest wavelength $\approx
10.5$\,\AA\ in \ab{fig:8-14MK}) are weak even in the
Fe-dominated model and completely absent in the Fe-solar model (features
at $\approx 13$\,\AA\ in \ab{fig:nature} at $z=0.24$). Hence, at
$z=0.35$ there is no absorption feature at 13\,\AA\ that can reproduce the
observations. With increasing \Teff\ the \ion{Fe}{xxv} lines become
slightly stronger, reaching maximum strength at \Teffw{12}, although, they
remain far weaker than the  \ion{Fe}{xxiv} $n=2-3$ absorption lines
\sA{fig:8-14MK}. The reason is the following. The \ion{Fe}{xxv}
ionisation stage dominates the considered atmospheres
(\ab{fig:structure}), but the lines at 10.33, 10.48\,\AA\ (for which
identifications were  suggested by CPM02) correspond to high energy ($E
\approx 6.7$ keV) levels for which the Boltzmann factor
$\exp (-E/kT)$ is low. Lines of \ion{Fe}{xxiv} originate in 
ground-level transitions that are more common than the high-energy level transitions of
\ion{Fe}{xxv}. We remark that occupation probabilities are
lower for high-energy levels (see e.g\@. \ab{fig:FE24_HM}). Therefore, the
high-energy levels of \ion{Fe}{xxv} are even less populated than the
ground level of \ion{Fe}{xxiv}.

Relative flux spectra for LTE models with \Teffw{10} are shown 
in \ab{fig:sul_f5}. This figure illustrates the influence of the chemical composition on the 
emergent flux. Clearly, the above conclusion about line identification 
is unchanged.

Therefore, from our results it is more likely that the observed features
are due to the resonance multiplet ($n=2-3$) of Li-like iron
(\ion{Fe}{xxiv}). This would, however, require a significantly lower
redshift of $z=0.24$. We briefly discuss the implications of
this result for the neutron star parameters.

\aba{fig:eos} shows the allowed values for mass and
radius in \exo\ for redshifts
$z=0.24$ and $z=0.35$ compared to various theoretical mass-radius
relations. These relations represent typical relations for a stiff (BalbN2),
moderate (SLy4), soft (BPAL12), and strange matter (SS) equation of state. 
Details are given in \citet{Hae06}.
While $z=0.35$ corresponds to radii of $R=9-12$~km for a mass-range of
$M=1.4-1.8$~M$_\odot$, our redshift $z=0.24$ implies larger radii of $R=12-15$~km,
which corresponds to stiff equations-of-state and excludes mass-radius
relations based on exotic matter. This result agrees with the study of
\exo\ by \citet{ozel06} using additional observational constraints. 
With $z=0.35$ and using this author's formulae and input values one obtains
minimum values of mass and radius, such that $M \geq 1.9$~M$_\odot$ and $R \geq 12.5$~km. 
A reduction in redshift to $z=0.24$ would have a negligible
effect on their radius determination but their lower mass limit would be reduced to
$M > 1.48$\,M$_\odot$.

In observations of \exo\ bursts by
\citet{cpm2008}, no significant photospheric line features were
detected. \citet{cpm2008} discussed several possible reasons for
this non-detection. They found no conclusive evidence for a different photospheric
temperature. Our models indicate that the \Ion{Fe}{24} lines become weaker towards higher
\Teff\ and disappear at about \Teffw{14} \sA{fig:8-14MK}. Considering
the rotation of \exo, we would expect a featureless spectrum as
soon as the temperature exceeds about \Teffw{12}.

\section{Conclusions}
\label{sect:conclusions}

We have performed model-atmosphere calculations to describe the
X-ray spectra of thermal radiation from neutron stars. 
We have compared our computed spectra with X-ray burst spectra of the LMXB
\exo. We have been unable to confirm the line identification by CPM02 as being due to
subordinate transitions of H- and He-like iron. These line features were
too weak at any \Teff\ and iron content. Our models suggested that a more
likely identification was the resonance line of Li-like iron. As a
consequence, the measured line redshift was $z=0.24$ rather than
$z=0.35$. This implied a larger neutron star radius of
$R=12-15$\,km for the mass range $M=1.4-1.8$\,M$_\odot$.

We compared results from two entirely different model codes, a LTE and a NLTE code,
and concluded that NLTE effects were less important than uncertainties in the
chemical composition of the bursting neutron-star atmospheres.
Given the current state of observations, NLTE effects on
continuum shape and spectral line profiles are negligible. On the other hand, we
investigated the influence of the various chemical compositions and found that the 
derived conclusion concerning line identification does not depend on the  
chemical composition.

We investigated the relevance of Compton electron scattering and
found that it was unimportant for solar composition models with $T_{\rm eff}
< 15\,\mathrm{MK}$.

\begin{figure}[ht!]
  \resizebox{\hsize}{!}{\includegraphics{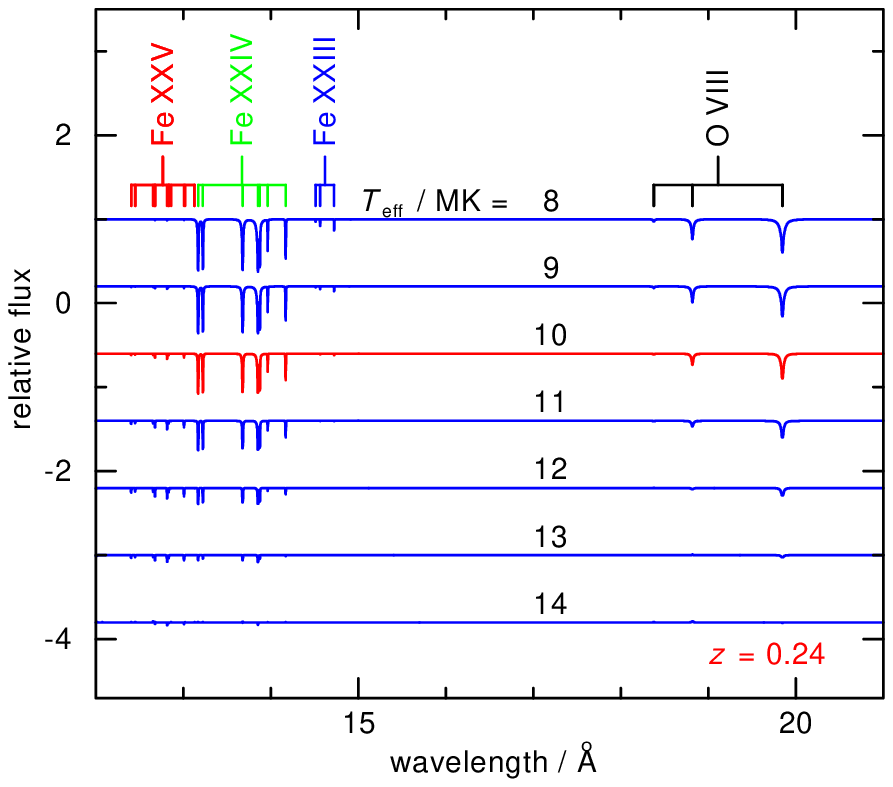}}
  \caption[]{Comparison of redshifted ($z=0.24$) NLTE model-atmosphere fluxes calculated from 
             \Teffw{8 - 14} models of solar iron abundance. Note that for clarity these 
             lines have not been convolved with either the instrument's resolution or
             any rotational profile (cf\@. \ab{fig:nature}).
            }
  \label{fig:8-14MK}
\end{figure}

The comparison of model-atmosphere spectra with blackbody flux distributions \sA{fig:10MK} has shown
that model-atmosphere spectra peak at higher energies and have a higher peak intensity.
A determination of \Teff\ by assuming blackbody spectra, as performed by CPM02, therefore overestimates \Teff\
\citep[cf\@.][]{rea2005}.

\begin{figure}[ht!]
  \resizebox{\hsize}{!}{\includegraphics{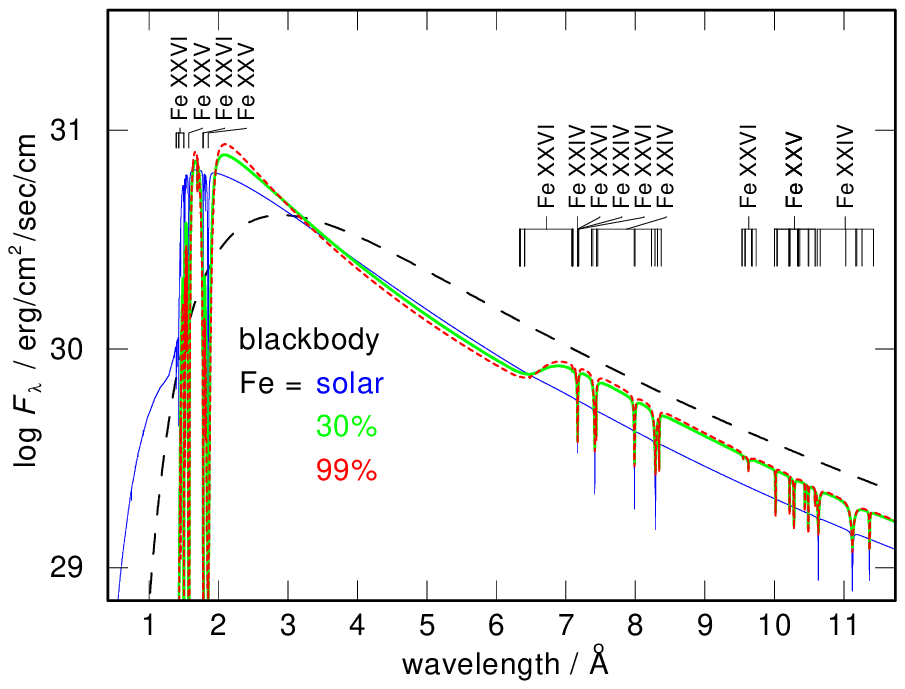}}
  \caption[]{Comparison of unredshifted NLTE model-atmosphere fluxes calculated from 
             \Teffw{10} models with different iron content (thin: solar, thick: 30\,\%
              Fe content, short dashes: 9\,\% Fe)
             The long-dashed line represents a blackbody spectrum.
            }
  \label{fig:10MK}
\end{figure}

\begin{figure}[ht!]
  \resizebox{\hsize}{!}{\includegraphics{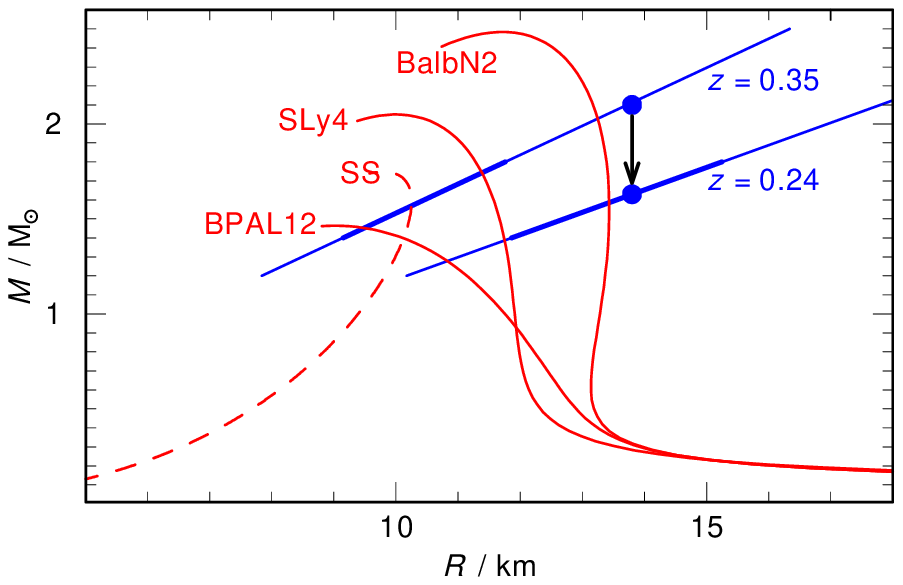}}
  \caption[]{Allowed values for $M$ and $R$ of \exo\ for redshifts $z=0.24$
and $z=0.35$ (straight lines; thick portions of the graphs denote the mass
range $1.4-1.8$~M$_\odot$) compared to various theoretical $M-R$ relations
\citep{Hae06}. The thick dot on the
$z=0.35$ line denotes the minimum $M$ and $R$ derived by \citet{ozel06}. The
arrow indicates the shift of this result when we assume $z=0.24$. A description of the
theoretical mass-radius relations is given in the text.}
  \label{fig:eos}
\end{figure}

\begin{acknowledgements}
We gratefully thank the referee for helpful comments that improved the final manuscript.
We thank Erik Kuulkers for pointing out an error in a previous version of the paper.
T.R\@. is supported by the \emph{German Astrophysical Virtual Observatory} (GAVO) project
of the German Federal Ministry of Education and Research (BMBF) under grant 05\,AC6VTB, 
V.S\@. by the DFG (grant We 1312/35$-$1 and SFB/TR7   
``Gravitational Wave Astronomy'') and partially supported by 
the Russian President's program for support of leading \mbox{science} schools (grant Nsh$-$4224.2008.2).
\end{acknowledgements}

\bibliographystyle{aa}
\bibliography{xmodels.bbl}

\begin{thebibliography}
\expandafter\ifx\csname natexlab\endcsname\relax\def\natexlab#1{#1}\fi

\bibitem[Badnell et al\@.(2005)Badnell et al\@.]{OP}
        Badnell, N\@. R., Bautista, M\@. A., Butler, K., et al\@.
        2005,
        MNRAS, 360, 458

\bibitem[Chang et al\@.(2006)Chang et al\@.]{cea2006}
        Chang, P., Morsink, S., Bildsten, L., \& Wasserman, I\@.
        2006,
        \apj, 636, L\,117
        
\bibitem[Chang et al\@.(2005)Chang, Bildsten \& Wasserman]{cbw2005}
        Chang, P., Bildsten, L., \& Wasserman, I\@.
        2005,
        \apj, 629, 998

\bibitem[Cottam et al\@.(2002)Cottam, Paerels \& M\'endez]{cpm2002}
        Cottam, J., Paerels, F., \& M\'endez, M\@. 
        2002 (CPM02),
        \nat, 420, 51

\bibitem[Cottam et al\@.(2008)Cottam et al\@.]{cpm2008}
        Cottam, J., Paerels, F., M\'endez, M., el al\@. 
        2008,
        \apj, 672, 504 

\bibitem[Cowley(1971)Cowley]{c71}
        Cowley, C\@. R\@. 
        1971, 
        The Observatory, 91, 139

\bibitem[Dere et al\@.(1997)Dere et al\@.]{dere97}
        Dere, K. P., Landi, E., Mason, H. E., Monsignori Fossi, B. C., \& Young, P. R\@.
        1997,
        \apjs, 125, 149

\bibitem[Haensel et al\@.(2006)Haensel et al\@.]{Hae06} 
        Haensel, P., Potekhin, A\@. Y., \& Yakovlev, D\@. G\@.
        2006, 
        Neutron Stars 1. Equation of State and Structure, Springer (Berlin: Springer)

\bibitem[Hubeny et al\@.(1994)Hubeny, Hummer \& Lanz]{Lanz.Hub:94}
        Hubeny, I., Hummer, D., \& Lanz, T\@. 
        1994, 
        \aap, 282, 151

\bibitem[Hummer \& Mihalas(1988)Hummer \& Mihalas]{hm1988} 
        Hummer D.G., \& Mihalas\,D\@.
        1988, 
        \apj, 331, 794

\bibitem[Ibragimov et al\@.(2003)Ibragimov et al\@.]{Ibragimov.etal:03}
        Ibragimov, A\@. A., Suleimanov, V\@. F., Vikhlinin, A., \& Sakhibullin, N\@. A\@. 
        2003, 
        Astronomy Reports, 47, 186

\bibitem[Lapidus et al\@.(1986)Lapidus, Sunyaev, \& Titarchuk]{Lapidusetal:86}
        Lapidus, I\@. I., Sunyaev, R\@. A., \& Titarchuk, L\@. G\@. 
        1986, 
        \sovast\ Lett\@., 12, 383

\bibitem[Lewin et al\@.(1993)Lewin et al\@.]{lpt93}
        Lewin, W\@. H\@. G., van Paradijs, J\@., \& Taam, R. E\@.
        1993,
        \ssr, 62, 223

\bibitem[London at al\@.(1986)London, Taam, \& Howard]{Londonetal:86}
        London, R\@. A., Taam, R\@. E., \& Howard, W\@. M\@. 
        1986, 
        \apj, 306, 170

\bibitem[Madej(1991)Madej]{Madej:91}
        Madej, J\@. 
        1991, 
        \apj, 376, 161

\bibitem[Madej et al\@.(2004)Madej, Joss, \& R{\'o}{\.z}a{\'n}ska]{Madej.etal:04}
        Madej, J., Joss, P\@. C., \& R{\'o}{\.z}a{\'n}ska, A\@. 
        2004, 
        \apj, 602, 904

\bibitem[Molkov et al\@.(2000)Molkov, Grebenev, \& Lutovinov]{mol00}
        Molkov, S\@. V., Grebenev, S\@. A., \& Lutovinov, A\@. A\@.  
        2000, 
        \aap, 357, L41

\bibitem[\"Ozel(2006)\"Ozel]{ozel06} 
        \"Ozel, F\@.
        2006,
        \nat, 441, 1115

\bibitem[Rauch(2003)Rauch]{r2003}
        Rauch, T\@.
        2003,
        \aap, 403, 709

\bibitem[Rauch \& Deetjen(2003)Rauch \& Deetjen]{rd2003} 
        Rauch, T., \& Deetjen, J\@. L\@. 
        2003, 
        in: {\it Stellar Atmosphere Modeling},
        eds\@. I\@. Hubeny, D\@. Mihalas, K\@. Werner, 
        The ASP Conference Series Vol\@. 288 (San Francisco: ASP), p\@. 103

\bibitem[Rauch et al\@.(2005)Rauch et al\@.]{rea2005} 
        Rauch, T., Orio M., Gonzales-Riestra, C., \& Still, M\@.
        2005,
        in: {\it 14$^\mathrm{th}$ European Workshop on White Dwarfs}, 
        eds\@. D\@. Koester, S\@. Moehler, 
        The ASP Conference Series Vol\@. 334 (San Francisco: ASP), p\@. 423

\bibitem[Rauch et al\@.(2007)Rauch et al\@.]{rea2007}
        Rauch, T., Ziegler, M., Werner, K., et al\@.
        2007,
        \aap, 470, 317 

\bibitem[Rogers \& Iglesias(1992)Rogers \& Iglesias]{OPAL}
        Rogers, F\@. J., \& Iglesias, C\@. A\@.
        1992,
        \aaps, 79, 507 

\bibitem[Suleimanov \& Poutanen(2006)Suleimanov \& Poutanen]{Sul.Pout:06}
        Suleimanov, V., \& Poutanen, J\@. 
        2006, 
        \mnras, 369, 2036
   
\bibitem[Suleimanov \& Werner(2007)Suleimanov \& Werner]{sw2007}
        Suleimanov, V., \& Werner, K\@.
        2007,
        \aap, 466, 661

\bibitem[Uns\"old(1968)Uns\"old]{u68}
        Uns\"old, A\@. 
        1968, 
        Physik der Sternatmosph\"aren, 
        Springer (Berlin: Springer)

\bibitem[van Dien(1949)van Dien]{vd49}
        van Dien, E\@. 
        1949, 
        \apj, 109, 452

\bibitem[Villarreal \& Strohmayer(2004)Villareal \& Strohmayer]{vs2004}
        Villarreal, A\@. R., \& Strohmayer, T\@. E\@.
        2004,
        \apj, 614, L121

\bibitem[Werner et al\@.(1991)Werner, Heber \& Hunger]{whh91}
        Werner, K., Heber, U., \& Hunger, K\@. 
        1991, 
        \aap, 244, 437

\bibitem[Werner et al\@.(2003)Werner et al\@.]{wea2003}  
        Werner, K., Dreizler, S., Deetjen, J\@. L., et al\@. 
        2003,
        in: {\it Stellar Atmosphere Modeling}, 
        eds\@. I\@. Hubeny, D\@. Mihalas, K\@. Werner, 
        The ASP Conference Series Vol\@. 288 (San Francisco: ASP), p\@. 31

\bibitem[Werner et al\@.(2007)Werner et al\@.]{wea2007}
        Werner, K., Nagel, T., Rauch, T., \& Suleimanov, V\@. 
        2007, 
        Adv\@. Space Res., 40, 1512

\bibitem[Werner et al\@.(2008)Werner, Rauch \& Kruk]{wrk2008}
        Werner, K., Rauch, T., \& Kruk, J\@. W\@. 
        2008, 
        \aap, 474, 591

\end{thebibliography}

\end{document}